\documentclass[prl,twocolumn,superscriptaddress]{revtex4}

\usepackage[ansinew]{inputenc}
\usepackage{graphicx}
\usepackage{amsmath}
\usepackage{amsthm}
\usepackage{layout}
\usepackage{float}
\usepackage{subfigure}
\usepackage{amsfonts}
\usepackage{amssymb}
\usepackage{txfonts}

\begin{document}

\title{Experimental violation of a Bell inequality with two different degrees of freedom of entangled particle pairs}
\author{Xiao-song Ma}
\affiliation{Institute for Quantum Optics and Quantum Information (IQOQI), Austrian
Academy of Sciences, Boltzmanngasse 3, A-1090 Vienna, Austria}
\affiliation{Faculty of Physics, University of Vienna, Boltzmanngasse 5, A-1090 Vienna,
Austria}
\author{Angie Qarry}
\affiliation{Institute for Quantum Optics and Quantum Information (IQOQI), Austrian
Academy of Sciences, Boltzmanngasse 3, A-1090 Vienna, Austria}
\author{Johannes Kofler}
\affiliation{Institute for Quantum Optics and Quantum Information (IQOQI), Austrian
Academy of Sciences, Boltzmanngasse 3, A-1090 Vienna, Austria}
\affiliation{Faculty of Physics, University of Vienna, Boltzmanngasse 5, A-1090 Vienna,
Austria}
\author{Thomas Jennewein}
\affiliation{Institute for Quantum Optics and Quantum Information (IQOQI), Austrian
Academy of Sciences, Boltzmanngasse 3, A-1090 Vienna, Austria}
\author{Anton Zeilinger}
\affiliation{Institute for Quantum Optics and Quantum Information (IQOQI), Austrian
Academy of Sciences, Boltzmanngasse 3, A-1090 Vienna, Austria}
\affiliation{Faculty of Physics, University of Vienna, Boltzmanngasse 5, A-1090 Vienna,
Austria}

\begin{abstract}
We demonstrate hybrid entanglement of photon pairs via the
experimental violation of a Bell inequality with two different
degrees of freedom (DOF), namely the path (linear momentum) of one
photon and the polarization of the other photon. Hybrid entangled
photon pairs are created by Spontaneous Parametric Down Conversion
and coherent polarization to path conversion for one photon. For
that photon, path superposition is analyzed, and polarization
superposition for its twin photon. The correlations between these
two measurements give an \textit{S}-parameter of $S=2.653\pm0.027$
in a CHSH inequality and thus violate local realism for two
different DOF by more than 24 standard deviations. This
experimentally supports the idea that entanglement is a fundamental
concept which is indifferent to the specific physical realization of
Hilbert space.
\end{abstract}

\maketitle

The assumption of local realism led Einstein, Podolsky, and Rosen
(EPR) to argue that quantum mechanics is not a complete theory
\cite{Einstein:1935}. In 1951 Bohm \cite{Bohm:1951} discussed a
system of two spatially separated and entangled spin-$\frac{1}{2}$
particles in order to illustrate the essential features of the EPR
paradox. All hidden-variable theories based on the joint assumptions
of locality and realism are at variance with the predictions of
quantum physics, as shown by the violation of the famous Bell
inequalities \cite{Bell:1964kc} using entangled spin-$\frac{1}{2}$
particles. Since the formulation of the Bell inequalities and later
of the Clauser-Horne-Shimony-Holt (CHSH) inequality
\cite{Clauser:1969}, numerous experiments based on
polarization-entangled photons have been performed that verified the
quantum-mechanical predictions \cite{Freedman:1972, Aspect:1982,
Weihs:1998}. Besides the polarization of photons, there are
theoretical proposals to test the Bell's inequality with the other
degrees of freedom (DOF), e.g. using the momentum \cite{Horne:1985,
Horne:1989, Zukowski:1988} or the emission time \cite{Franson:1989}
of entangled photon pairs. The experimental violation of Bell's
inequality based on the momentum and phase was demonstrated by
Rarity and Tapster \cite{Rarity:1990}, while the time-bin
entanglement has been employed in the fiber-based quantum
cryptography and communication \cite{Gisin:2002}.

Here we follow the proposal in \cite{Zukowski:1991} and experimentally
realize \textit{hybrid} entanglement, which is the entanglement between
\textit{different} degrees of freedom of a particle \textit{pair}. We
specifically demonstrate the hybrid entanglement between the polarization of
a photon from a photon pair and the path (momentum) of its twin. We want to
stress that hybrid entanglement is in principle different from so-called
hyper-entanglement \cite{Barreiro:2005}. A hyper-entangled state is a tensor
product of entangled states in each \textit{individual} DOF. Therefore,
there is no entanglement between different DOF. \textit{A hybrid entangled
state cannot be factorized into states of individual DOF only.} In a
hyper-entangled state of, say, two particles joint properties of the same
degree of freedom are well defined at the expense of defining individual
properties. The joint properties allow to make predictions for experimental
situations where both particles are measured in one and the same degree of
freedom. With hybrid entanglement the situation is different. There, the
defined joint properties are such that they link one degree of freedom of
one particle with another degree of freedom of the other particle, where
those degrees may even be defined in Hilbert spaces of different
dimensionalities as, e.g., polarization and linear momentum. While the
Hilbert space structure of quantum mechanics demands the existence of such
hybrid-entangled states, they have not been shown experimentally until now.

The entanglement between the polarization and the momentum DOF \cite%
{Boschi:1998, Michler:2000} as well as between the polarization and
the orbital angular momentum DOF \cite{Barreiro:2008} of a
\textit{single} photon, and between the spatial and spin DOF of a
\textit{single} neutron \cite{Hasegawa:2003} was demonstrated
experimentally. The idea to convert the polarization entanglement to
path entanglement of a photon pair was realized in
\cite{Strekalov:1996}. There have also been experimental
realizations of two-photon four-qubit cluster states with
entanglement between both path and polarization
\cite{Vallone:2007,Chen:2007}. On the other hand, entanglement
between the same degree of freedom of different physical systems has
also been realized. In many atom-photon experiments entanglement has
been demonstrated between the spin of the atom state and the spin
(i.e.\ polarization) of the photon \cite{Raimond:2001}.

In this letter, we demonstrate hybrid entanglement of photon pairs between
two different degrees of freedom, namely path (linear momentum) and
polarization, via the experimental violation of the CHSH inequality.
Normally, in the case of the polarization entanglement of a photon pair, the
maximum violation of the CHSH inequality is established with the polarizers
oriented at ${(-22.5^{\circ}, 22.5^{\circ})}$ at Bob's side and ${%
(0^{\circ}, 45^{\circ})}$ at Alice's side, while in the case of path
entanglement it is established with the phase shift at ${(-45^{\circ},
45^{\circ})}$ at Bob's side and ${(0^{\circ}, 90^{\circ})}$ at Alice's side.
In order to maximally violate the CHSH inequality for the hybrid
entanglement, the polarizer at Bob's side (photon B) is oriented at the
\textit{angles} of ${(-22.5^{\circ}, 22.5^{\circ})}$ and the phase shifter
at Alice's side (photon A) is adjusted at the \textit{phase} of ${%
(0^{\circ}, 90^{\circ})}$. This manifests the hybrid nature of our entangled
photon pairs. A very important feature of the present experiment is, that
the interferometer---the analyzer of the path DOF---is calibrated strictly
locally and before the correlation measurements. Therefore, it is possible
to independently choose the settings on each side. We also applied this
specific hybrid entanglement to perform a quantum eraser experiment under
strict (Einstein) locality condition, which will be presented elsewhere.

The scheme of our experiment is shown in FIG. 1. First, we create
the polarization-entangled EPR-Bell state $|\Phi ^{+}\rangle =\frac{1}{\sqrt{2}}%
(\left\vert H\right\rangle _{\text{A}}\left\vert V\right\rangle _{\text{B}%
}+\left\vert V\right\rangle _{\text{A}}\left\vert H\right\rangle _{\text{B}%
}) $, where $\left\vert H\right\rangle $ and $\left\vert V\right\rangle $
denote the horizontal and vertical linearly polarized quantum states
respectively, and A and B index the spatial modes of the photons.
\begin{figure*}[t]
\centerline{\includegraphics[width=0.75\textwidth]{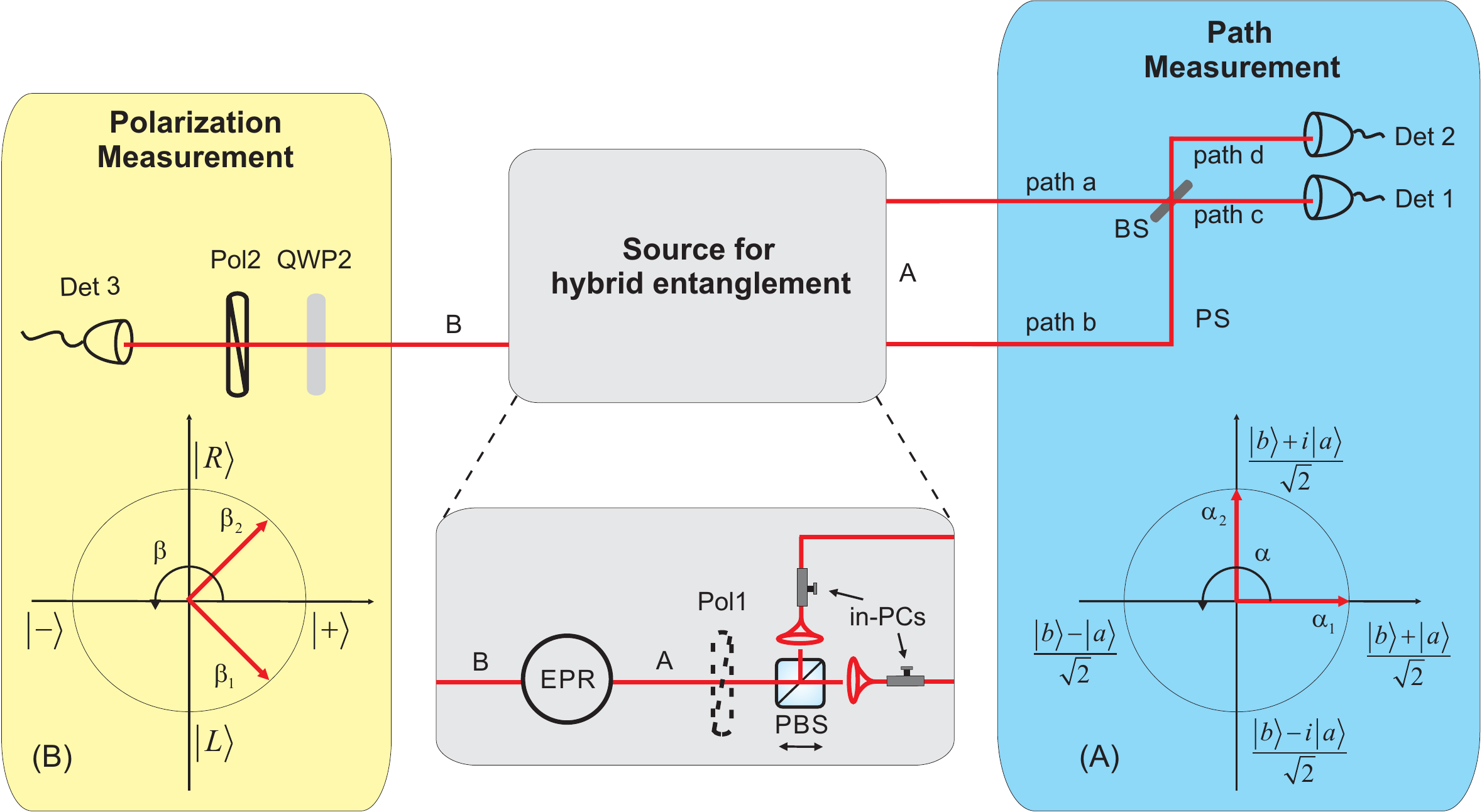}}
\caption{(Color online.) Experimental Setup. Polarization-entangled
photon pairs are generated in an EPR-Bell state via spontaneous
parametric down conversion (SPDC). A picosecond-pulsed Nd:Vanadate
laser emitting light at
the wavelength of $355$~nm after frequency tripling (repetition rate of $76$%
~MHz and average power of $200$~mW) pumps a $\protect\beta $-barium borate ($%
\protect\beta $-BBO) crystal in a cross rings type-II scheme of SPDC
\protect\cite{Kwiat:1995}. Good spectrum and spatial mode
overlapping is achieved by using interference filters with $1$~nm
bandwidth centered around $710$~nm and by collecting the entangled
photon-pairs into single mode fibers \protect\cite{Kurtsiefer:2001}.
In order to create the hybrid entangled state (\protect\ref{qs3}),
the source also consists of a polarizing beam splitter (PBS), two
in-line polarization controllers (in-PCs) and an additional linear
polarizer (Pol1) [see main text for details]. The photon in spatial
mode A is directed towards the interferometric path measurement
setup. We combine both paths on a single-mode fiber beam splitter
(BS) and the length of the whole interferometer is about 2~m. The
phase scanner (PS) is realized via position change ($x$) of the PBS.
The photon in spatial mode B is directed towards the polarization
measurement setup. It consists of a quarter waveplate (QWP2) and a
linear polarizer (Pol2) with the transmission axis oriented along
angle $\protect\phi $, which together allow to project photon B into
the desired polarization states. Both photon A and photon B are
detected by multimode fiber coupled silicon avalanche photodiodes
(Det 1, 2,
3). Photon A is analyzed in the superposition of the two path states along $%
\protect\alpha _{1}$, $\protect\alpha _{2}$ and their orthogonal directions
on its Bloch sphere shown in the inset (A). Photon B is analyzed in the
superposition of the polarization states along directions $\protect\beta %
_{1} $, $\protect\beta _{2}$ and their orthogonal directions on its Bloch
sphere shown in the inset (B).}
\label{fig1}
\end{figure*}
Next we will investigate how the quantum state $|\Phi ^{+}\rangle $ evolved
in the setup. A polarizing beam splitter (PBS) transmits the horizontal and
reflects the vertical polarization state of Photon A. Thus, the PBS acts as
a deterministic polarization-momentum converter. Two in-line polarization
controllers (in-PCs) are used to rotate the orthogonal polarization states
of photon A in path a and b ($\left\vert H\right\rangle _{\text{Aa}}$ and $%
\left\vert V\right\rangle _{\text{Ab}}$ ) to an identical one ($\left\vert
\theta ,\gamma \right\rangle _{\text{Aa}}$ and $\left\vert \theta ,\gamma
\right\rangle _{\text{Ab}}$, where $\left\vert \theta ,\gamma \right\rangle
_{\text{Aa}}=\left\vert \theta ,\gamma \right\rangle _{\text{Ab}}=\cos {%
\theta }\left\vert H\right\rangle +\exp (i\gamma )\sin {\theta }\left\vert
V\right\rangle $) and thus eliminate the polarization distinguishability of
the two paths. Hence from now on we will ignore the polarization of photon A
and label it with its path quantum states, where $\left\vert a\right\rangle
_{\text{A}}\equiv \left\vert \theta ,\gamma \right\rangle _{\text{Aa}}$ and $%
\left\vert b\right\rangle _{\text{A}}\equiv \left\vert \theta
,\gamma \right\rangle _{\text{Ab}}$. Therefore, the source creates
the hybrid entangled state between the path of photon A and the
polarization of photon B:%
\begin{equation}
|\Phi _{\text{hybrid}}^{+}\rangle =\frac{1}{\sqrt{2}}(\left\vert
b\right\rangle _{\text{A}}\left\vert V\right\rangle _{\text{B}}+\left\vert
a\right\rangle _{\text{A}}\left\vert H\right\rangle _{\text{B}}),
\label{qs3}
\end{equation}%
The superposition states of the two paths of photon A are varied and
analyzed by a modified Mach-Zehnder interferometer. After a phase scanner
(PS) and beam splitter (BS), the state becomes $|\Phi _{\text{hybrid}%
}^{+\prime }\rangle =\frac{1}{2}[(\left\vert d\right\rangle _{\text{A}}+%
\mathrm{i}\left\vert c\right\rangle _{\text{A}})\left\vert V\right\rangle _{%
\text{B}}+\exp {(\mathrm{i}\alpha )}(\left\vert c\right\rangle _{\text{A}}+%
\mathrm{i}\left\vert d\right\rangle _{\text{A}})\left\vert H\right\rangle _{%
\text{B}}]$. On photon B's side, by proper polarization projection, the
hybrid entangled state becomes:
\begin{eqnarray}
|\Phi _{\text{hybrid}}^{+\prime \prime }\rangle \; &=&\;\frac{1}{2}(\mathrm{e%
}^{\mathrm{i\kappa _{1}}}\sqrt{1+\sin {(\alpha +\beta )}}\left\vert
c\right\rangle _{\text{A}}\left\vert \beta \right\rangle _{\text{B}})  \notag
\\
&&+\ \mathrm{e}^{\mathrm{i\kappa _{2}}}\sqrt{1-\sin {(\alpha +\beta )}}%
\left\vert d\right\rangle _{\text{A}}\left\vert \beta \right\rangle _{\text{B%
}}  \notag \\
&&+\ \mathrm{e}^{\mathrm{i\kappa _{3}}}\sqrt{1-\sin {(\alpha +\beta )}}%
\left\vert c\right\rangle _{\text{A}}\left\vert \beta ^{\bot }\right\rangle
_{\text{B}}  \notag \\
&&+\ \mathrm{e}^{\mathrm{i\kappa _{4}}}\sqrt{1+\sin {(\alpha +\beta )}}%
\left\vert d\right\rangle _{\text{A}}\left\vert \beta ^{\bot }\right\rangle
_{\text{B}}).  \label{qs2}
\end{eqnarray}%
Here, $\left\vert \beta \right\rangle _{\text{B}}=\frac{1}{\sqrt{2}}%
(\left\vert H\right\rangle _{\text{B}}+\exp {(\mathrm{i}\beta )}\left\vert
V\right\rangle _{\text{B}})$ and $\left\vert \beta ^{\bot }\right\rangle _{%
\text{B}}=\frac{1}{\sqrt{2}}(\left\vert H\right\rangle _{\text{B}}-\exp {(%
\mathrm{i}\beta )}\left\vert V\right\rangle _{\text{B}})$ respectively, $%
\left\vert c\right\rangle _{\text{A}}$ and $\left\vert d\right\rangle _{%
\text{A}}$ are the spatial modes after the BS, and $\mathrm{\kappa _{1}}$, $%
\mathrm{\kappa _{2}}$, $\mathrm{\kappa _{3}}$ and $\mathrm{\kappa _{4}}$ are
the phases of the four different coincidence terms which are not important
in the present experiment.

On the photon A side, we tune the local phase difference between the two
path quantum states ($\left\vert a\right\rangle_{\text{A}}$ and $\left\vert
b\right\rangle_{\text{A}}$), which corresponds to the phase $\alpha$ of the
interferometer in (\ref{qs2}). Scanning this phase $\alpha$ with PS to ${%
(\alpha_{1}\equiv0^{\circ}, \alpha_{2}\equiv90^{\circ},
\alpha_{1}^{\bot}\equiv180^{\circ}, \alpha_{2}^{\bot}\equiv-90^{\circ})}$
and detecting the photon with Det1 and Det2 is like projecting the path
states of photon A into the states $\left\vert \alpha_{1}\right\rangle\equiv%
\frac{1}{\sqrt{2}}(\left\vert b\right\rangle+\left\vert a\right\rangle)$, $%
\left\vert
\alpha_{2}\right\rangle\equiv\frac{1}{\sqrt{2}}(\left\vert
b\right\rangle+\mathrm{i}\left\vert a\right\rangle)$, $\left\vert
\alpha_{1}\right\rangle^{\bot}\equiv\frac{1}{\sqrt{2}}(\left\vert
b\right\rangle-\left\vert a\right\rangle)$ and $\left\vert
\alpha_{2}\right\rangle^{\bot}\equiv\frac{1}{\sqrt{2}}(\left\vert
b\right\rangle-\mathrm{i}\left\vert a\right\rangle)$ respectively,
as shown in the inset of FIG. 1(A). The relation between the
position $x$ of the PBS and the phase of the interferometer $\alpha$
is $x=\frac{\alpha\lambda}{2\pi} $. On the photon B side, we can
tune the phase between the two polarization quantum states
($\left\vert H\right\rangle_{\text{B}}$ and $\left\vert
V\right\rangle_{\text{B}}$), which corresponds to the phase $\beta$ in (\ref%
{qs2}). By setting the QWP2 at $-45^{\circ}$ oriented relative to the
horizontal direction and rotating Pol2 such that $\beta$ is equal to ${%
(\beta_{1}\equiv-45^{\circ}, \beta_{2}\equiv45^{\circ},
\beta_{1}^{\bot}\equiv135^{\circ}, \beta_{2}^{\bot}\equiv-135^{\circ})}$, we
are able to project the polarization states of photon B into the desired
states $\left\vert \beta_{1}\right\rangle=\frac{1}{\sqrt{2}}(\left\vert
H\right\rangle_{\text{B}}+\frac{1}{\sqrt{2}}(1-\mathrm{i})\left\vert
V\right\rangle_{\text{B}})$, $\left\vert \beta_{2}\right\rangle=\frac{1}{%
\sqrt{2}}(\left\vert H\right\rangle_{\text{B}}+\frac{1}{\sqrt{2}}(1+\mathrm{i%
})\left\vert V\right\rangle_{\text{B}})$, $\left\vert
\beta_{1}\right\rangle^{\bot}=\frac{1}{\sqrt{2}}(\left\vert H\right\rangle_{%
\text{B}}+\frac{1}{\sqrt{2}}(-1+\mathrm{i})\left\vert V\right\rangle_{\text{B%
}})$ and $\left\vert {\beta}_{2}\right\rangle^{\bot}=\frac{1}{\sqrt{2}}%
(\left\vert H\right\rangle_{\text{B}}+\frac{1}{\sqrt{2}}(-1-\mathrm{i}%
)\left\vert V\right\rangle_{\text{B}})$ respectively, as shown in the inset
(B) of FIG. 1. The relation between the orientation angle of Pol2 and $%
\beta$ is $\phi=-\frac{\beta}{2}$.
\begin{figure}[t]
\centerline{\includegraphics[width=0.45\textwidth]{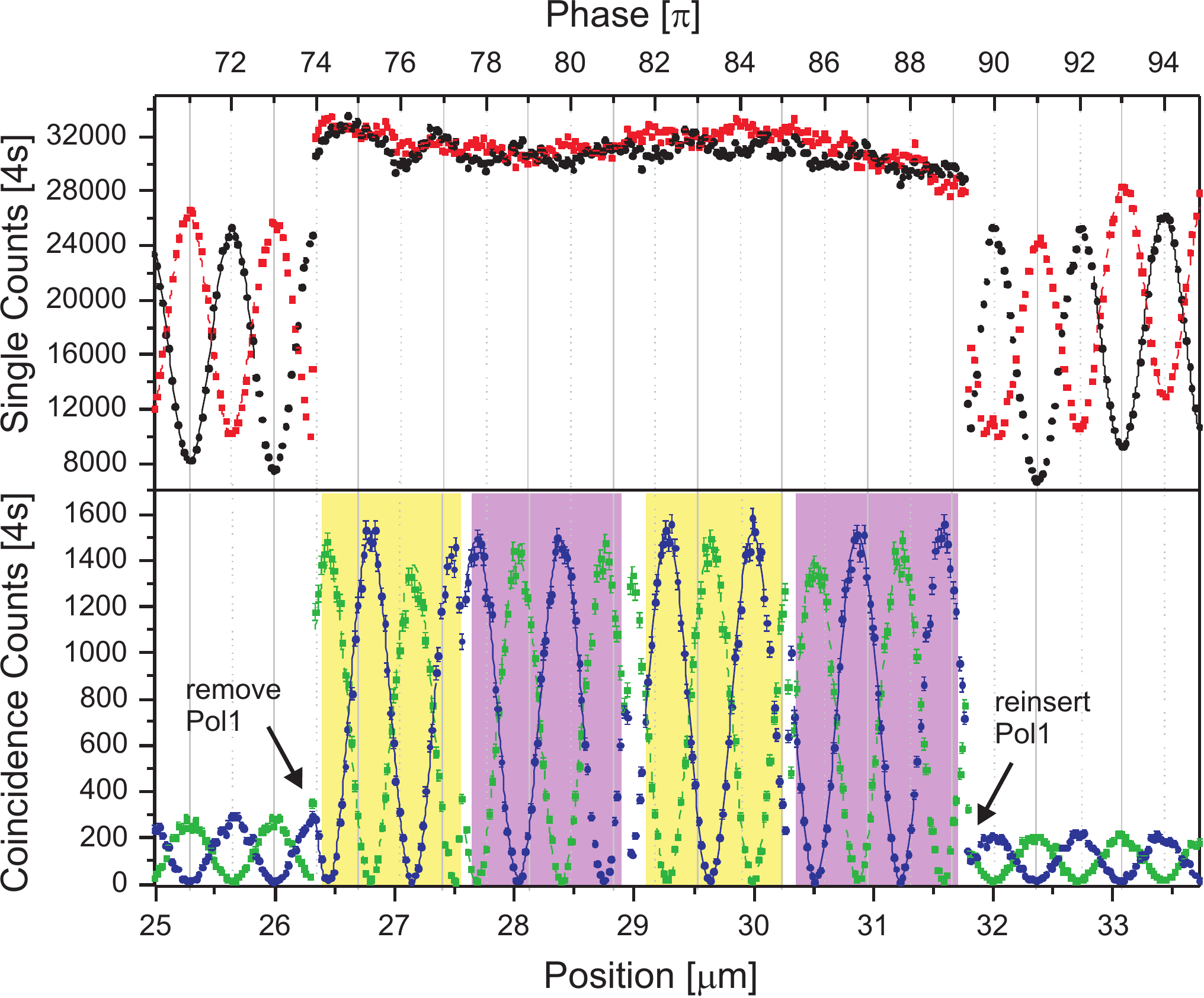}}
\caption{(Color online.) Experimental results. a. The single counts
of Det1 (red square dots) and Det2 (black circular dots) are fitted
with sinusoidal curves (red dash and black solid lines for Det1 and
Det2 respectively) at the beginning and the end in order to
calibrate the local phase of the Mach-Zehnder interferometer. b. The
coincidence counts between Det1 and Det3 (green square dots) and
Det2 with Det3 (blue circular dots) and the corresponding sinusoidal
fits (green dash and blue solid lines respectively). They are used
to construct the correlation coefficients in order to violate the
Bell inequality. Alternating color (grey) shadings are designating
the different settings of Pol2. The actions of removing and
reinserting Pol1 are identified with arrows. The error bars are the
square roots of the corresponding counts. } \label{fig2}
\end{figure}

Experimentally, we measured in three steps: (I) we inserted Pol1 oriented at
$45^{\circ}$ into the setup. Then the entanglement is erased and photon A is
in a coherent superposition of taking path a ($\left\vert a\right\rangle_{%
\text{A}}$) or path b ($\left\vert b\right\rangle_{\text{A}}$). In
FIG. 2a, we show the single counts of Det1 (red square dots) and
Det2 (black circular dots). Two oppositely modulated data curves, as
a function of the relative phase change of the two paths, enable us
to find the absolute value of the local phase of the interferometer.
We define $\alpha\equiv2n\pi$ ($n$ is an integer) when Det2 has
maximum counts. Thus, the coincidence counts of Det1 with Det3
(green square dots in FIG. 2) and Det2 with Det3 (blue circular dots
in FIG. 2) are oscillating in phase with the corresponding single
counts.

(II) We remove Pol1 and measure the coincidence counts of Det1 with
Det3 and Det2 with Det3. From these coincidence counts we construct
the correlation coefficients for the violation of the Bell
inequality. When we take out Pol1, there are two important features
in FIG. 2. First, the oscillations of single counts ceased and this
can be explained by Equation (\ref{qs2}). For instance, one can
calculate the probability amplitude for $\left\vert
c\right\rangle_{\text{A}}$, which is a sum of two oppositely
modulated sinusoidal functions. Thus, the single counts of Det1 are
insensitive to the phase change both "locally" ($\alpha$) and
"nonlocally" ($\beta$). The same reasoning applies to the single
counts of Det2 as well. Second, the coincidence counts behave
differently relative to the single counts. The
coincidence counts keep oscillating as we are scanning the local phase ($%
\alpha$) and the oscillating amplitude increases. The reason for the
increase is that we first aligned Pol2 at $-22.5^{\circ}$ and Pol1 at $%
-45^{\circ}$ and theoretically the corresponding coincidence counts are only
$0.146$ of the coincidence counts of the case when Pol1 is removed.
Experimentally we found that was about $0.19$. Moreover, there is a phase
jump between the oscillating curves of the coincidence counts of the two
cases with or without Pol1. For example, the coincidence counts between Det1
and Det3 are proportional to the joint probability for detecting photon A in
path c ($\left\vert c\right\rangle_{\text{A}}$) and detecting the
polarization of photon B along $\beta$, which is proportional to $1+\sin{%
(\alpha)}$ with Pol1 and proportional to $1+\sin{(\alpha+\beta)}$ without
Pol1. Experimentally, as stated above, we first align Pol2 at $-22.5^{\circ}$
and Pol1 at $-45^{\circ}$, which corresponds to a phase difference of $%
225^{\circ}$. The measured value is $230^{\circ}$. This allows to
quantitatively explain that the coincidence counts are expected to be $0.18$
of the coincidence of the case when Pol1 is removed. Then we scan the local
phase continuously and set the orientation angle of Pol2 to ${%
(-22.5^{\circ}, 22.5^{\circ}, 67.5^{\circ}, 112.5^{\circ})}$
sequentially, which corresponds to $(\beta_{2}, \beta_{1},
\beta_{2}^{\bot}, \beta_{1}^{\bot})$. These four different settings
are designated with four alternated color (grey) shaded regions in
FIG. 2b. Due to the reasons stated above, there are phase jumps of
the coincidence counts between the different settings of Pol2. The
phase jumps between the neighboring regions are
expected to be $90^{\circ}$, while $89.2^{\circ}$, $92.4^{\circ}$ and $%
86.8^{\circ}$ were the measured values, respectively. These four regions of
the data are enough to construct the correlation coefficients and to violate
the Bell inequality.

(III) After we get the coincidence data, we insert Pol1 back again to
determine the phase drift during the whole measurement cycle. We get a $%
2.0^{\circ}$ phase difference on average. Without subtracting the
accidental coincidence counts, the interference visibilities of the
coincidence counts are above 96\% for all four settings. The
wavelength of all the fits (including single counts and coincidence
counts) is fixed to 708.6 nm.

\begin{figure}[t]
\centerline{\includegraphics[width=0.45\textwidth]{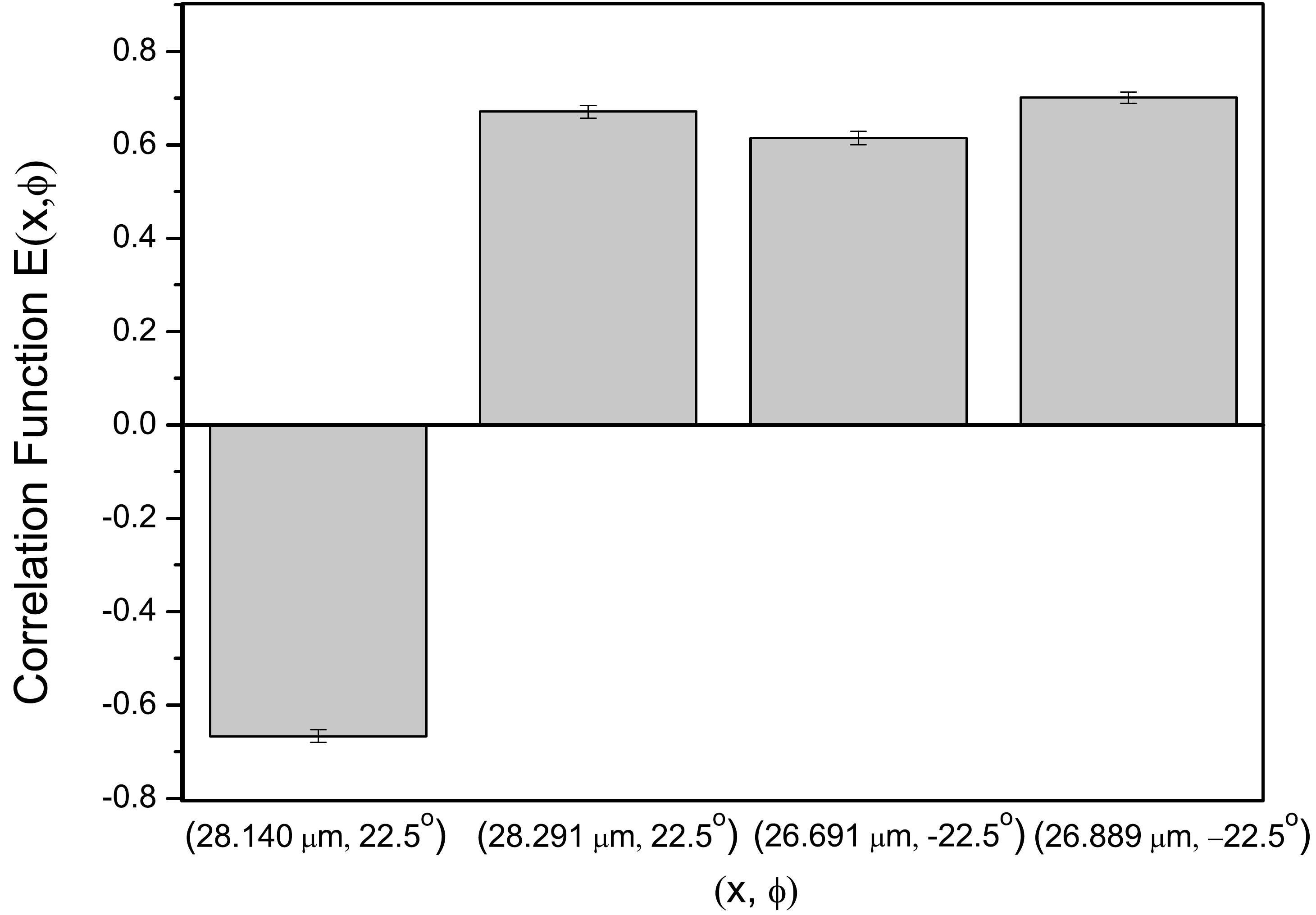}}
\caption{Four correlation functions of the CHSH inequality for four
different settings. Operationally, the setting on Photon A side is
given by the position of the phase scanner $x$ and on Photon B side
it is the orientation of the polarizer $\protect\phi$. This
manifests the hybrid nature of the entangled photon pair. The error
bars represent its statistical errors. } \label{fig3}
\end{figure}

Given a setting pair $(\alpha_{i}, \beta_{j})$, which are the orientations
of the vectors of the analyzers on the Bloch sphere of photon A and B
respectively, the correlation coefficients are defined as:
\begin{equation}
E(\alpha_{i}, \beta_{j})=\frac{C(\alpha_{i}, \beta_{j})+C(\alpha_{i}^{\bot},
\beta_{j}^{\bot})-C(\alpha_{i}^{\bot}, \beta_{j})-C(\alpha_{i},
\beta_{j}^{\bot})}{C(\alpha_{i}, \beta_{j})+C(\alpha_{i}^{\bot},
\beta_{j}^{\bot})+C(\alpha_{i}^{\bot}, \beta_{j})+C(\alpha_{i},
\beta_{j}^{\bot})},  \label{cc1}
\end{equation}
where $C(\alpha_{i}, \beta_{j})$ and $C(\alpha_{i}^{\bot}, \beta_{j})$ ($%
C(\alpha_{i}^{\bot}, \beta_{j}^{\bot})$ and $C(\alpha_{i}, \beta_{j}^{\bot})$%
) are the coincidence counts of Det1 with Det3 and Det2 with Det3
respectively, given the local phase of interferometer on photon A side is $%
\alpha_{i}$ ($\alpha_{i}^{\bot}$) and the orientation of polarizer on photon
B side is such that $\beta=\beta_{j}$ ($\beta_{j}^{\bot}$) with $i,j=1,2$.
From the state (\ref{qs2}), it follows that $E(\alpha_{i}, \beta_{j})=\sin({%
\alpha_{i}+\beta_{j})}$. If local realism is valid, such correlation
coefficients must satisfy the CHSH inequality:
\begin{equation}
S=-E(\alpha_{1}, \beta_{1})+E(\alpha_{1}, \beta_{2})+E(\alpha_{2},
\beta_{1})+E(\alpha_{2}, \beta_{2})\leq{2}  \label{BI}
\end{equation}
But quantum mechanics predicts values up to $2\sqrt{2}$.

The correlation coefficients are calculated from the data from FIG.
2b, which are $E(\alpha_{1}, \beta_{1})=E(28.140 \, \mu\text{m},
22.5^{\circ})=-0.666 \pm 0.014$, $E(\alpha_{1}, \beta_{2})=E(28.291 \, \mu%
\text{m}, 22.5^{\circ})=0.671\pm 0.014$, $E(\alpha_{2},
\beta_{1})=E(26.691 \, \mu\text{m}, -22.5^{\circ})=0.615 \pm 0.014$,
and $E(\alpha_{2}, \beta_{2})=E(26.889 \, \mu\text{m},
-22.5^{\circ})=0.701\pm 0.012$ respectively, as shown in FIG. 3. The
\textit{S}-parameter calculated from those four correlation
coefficients equals to $S=2.653 \pm 0.027$, which violates the
classical bound ($|S|=2$) by more than 24 standard deviations.

\textit{Conclusion}.---Hybrid entanglement is not only of fundamental
interest. It also could be useful in quantum information processing, e.g.\
the quantum repeater \cite{Duan:2001}. It is not limited to the case of path
(linear momentum) and polarization, as we have shown in this letter, but
also should be possible for other degrees of freedom, e.g. frequency,
orbital angular momentum etc. of photons.

\begin{acknowledgements}
We are grateful to R. Ursin, {\v C}. Brukner and M. \.{Z}ukowski for
discussions. We acknowledge support by the IST funded Integrated
Project QAP (Contract No.\ 015846) of the European Commission and
the Doctoral Program CoQuS of the Austrian Science Foundation (FWF).
\end{acknowledgements}


\begin{thebibliography}
\bibitem{} \expandafter\ifx\csname
natexlab\endcsname\relax

\fi \expandafter\ifx\csname bibnamefont\endcsname\relax

\fi \expandafter\ifx\csname bibfnamefont\endcsname\relax

\fi \expandafter\ifx\csname citenamefont\endcsname\relax

\fi \expandafter\ifx\csname url\endcsname\relax

\fi \expandafter\ifx\csname urlprefix\endcsname\relax

\fi \providecommand{\bibinfo}[2]{#2}
\providecommand{\eprint}[2][]{\url{#2}}

\bibitem[Einstein et~al.(1935)Einstein, Podolsky, and Rosen]{Einstein:1935} %
\bibinfo{author}{\bibfnamefont{A.}~\bibnamefont{Einstein}}, %
\bibinfo{author}{\bibfnamefont{B.}~\bibnamefont{Podolsky}}, and %
\bibinfo{author}{\bibfnamefont{N.}~\bibnamefont{Rosen}}, %
\bibinfo{journal}{Phys. Rev.} \textbf{\bibinfo{volume}{47}}, %
\bibinfo{pages}{777} (\bibinfo{year}{1935}).

\bibitem[Bohm(1951)]{Bohm:1951} \bibinfo{author}{\bibfnamefont{D.}~%
\bibnamefont{Bohm}}, \emph{\bibinfo{title}{Quantum Theory}} (%
\bibinfo{publisher}{Prentice-Hall}, \bibinfo{address}{Englewood Cliffs, NJ}, %
\bibinfo{year}{1951}).

\bibitem[Bell(1964)]{Bell:1964kc}
\bibinfo{author}{\bibfnamefont{J.~S.}
\bibnamefont{Bell}}, \bibinfo{journal}{Physics} \textbf{\bibinfo{volume}{1}}%
, \bibinfo{pages}{195} (\bibinfo{year}{1964}).

\bibitem[Clauser et~al.(1969)Clauser, Horne, Shimony, and Holt]%
{Clauser:1969} \bibinfo{author}{\bibfnamefont{J.~F.} \bibnamefont{Clauser}}, %
\bibinfo{author}{\bibfnamefont{M.~A.} \bibnamefont{Horne}}, %
\bibinfo{author}{\bibfnamefont{A.}~\bibnamefont{Shimony}}, and %
\bibinfo{author}{\bibfnamefont{R.~A.} \bibnamefont{Holt}}, %
\bibinfo{journal}{Phys. Rev. Lett.} \textbf{\bibinfo{volume}{23}}, %
\bibinfo{pages}{880} (\bibinfo{year}{1969}).

\bibitem[Freedman and Clauser(1972)]{Freedman:1972} \bibinfo{author}{%
\bibfnamefont{S.~J.} \bibnamefont{Freedman}} and \bibinfo{author}{%
\bibfnamefont{J.~F.} \bibnamefont{Clauser}},
\bibinfo{journal}{Phys. Rev.
Lett.} \textbf{\bibinfo{volume}{28}}, \bibinfo{pages}{938} (%
\bibinfo{year}{1972}).

\bibitem[Aspect et~al.(1982)Aspect, Grangier, and Roger]{Aspect:1982} %
\bibinfo{author}{\bibfnamefont{A.}~\bibnamefont{Aspect}}, %
\bibinfo{author}{\bibfnamefont{P.}~\bibnamefont{Grangier}}, and %
\bibinfo{author}{\bibfnamefont{G.}~\bibnamefont{Roger}}, %
\bibinfo{journal}{Phys. Rev. Lett.} \textbf{\bibinfo{volume}{49}}, %
\bibinfo{pages}{91} (\bibinfo{year}{1982}).

\bibitem[Weihs et~al.(1998)Weihs, Jennewein, Simon, Weinfurter, and Zeilinger%
]{Weihs:1998} \bibinfo{author}{\bibfnamefont{G.}~\bibnamefont{Weihs}}, %
\bibinfo{author}{\bibfnamefont{T.}~\bibnamefont{Jennewein}}, %
\bibinfo{author}{\bibfnamefont{C.}~\bibnamefont{Simon}}, \bibinfo{author}{%
\bibfnamefont{H.}~\bibnamefont{Weinfurter}}, and \bibinfo{author}{%
\bibfnamefont{A.}~\bibnamefont{Zeilinger}}, \bibinfo{journal}{Phys. Rev.
Lett.} \textbf{\bibinfo{volume}{81}}, \bibinfo{pages}{5039} (%
\bibinfo{year}{1998}).

\bibitem[Horne and Zeilinger(1985)]{Horne:1985} \bibinfo{author}{%
\bibfnamefont{M.~A.} \bibnamefont{Horne}} and \bibinfo{author}{%
\bibfnamefont{A.}~\bibnamefont{Zeilinger}}, \emph{%
\bibinfo{title}{Proceedings of the Symposium on the Foundations of
  Modern Physics}} (\bibinfo{publisher}{World Scientific, Singapore}, %
\bibinfo{year}{1985}).

\bibitem[Horne et~al.(1989)Horne, Shimony, and Zeilinger]{Horne:1989} %
\bibinfo{author}{\bibfnamefont{M.~A.} \bibnamefont{Horne}}, %
\bibinfo{author}{\bibfnamefont{A.}~\bibnamefont{Shimony}}, and %
\bibinfo{author}{\bibfnamefont{A.}~\bibnamefont{Zeilinger}}, %
\bibinfo{journal}{Phys. Rev. Lett.} \textbf{\bibinfo{volume}{62}}, %
\bibinfo{pages}{2209} (\bibinfo{year}{1989}).

\bibitem[Marek~Zukowski(1988)]{Zukowski:1988}
\bibinfo{author}{\bibfnamefont{M.}~\bibnamefont{\ifmmode~\dot{Z}\else
  \.{Z}\fi{}ukowski}} and
\bibinfo{author}{\bibfnamefont{J.}
\bibnamefont{Pykacz}}, \bibinfo{journal}{Phys. Lett. A} \textbf{%
\bibinfo{volume}{127}}, \bibinfo{pages}{1} (\bibinfo{year}{1988}).

\bibitem[Franson(1989)]{Franson:1989}
\bibinfo{author}{\bibfnamefont{J.~D.}
\bibnamefont{Franson}}, \bibinfo{journal}{Phys. Rev. Lett.} \textbf{%
\bibinfo{volume}{62}}, \bibinfo{pages}{2205} (\bibinfo{year}{1989}).

\bibitem[Rarity and Tapster(1990)]{Rarity:1990} \bibinfo{author}{%
\bibfnamefont{J.~G.} \bibnamefont{Rarity}} and \bibinfo{author}{%
\bibfnamefont{P.~R.} \bibnamefont{Tapster}},
\bibinfo{journal}{Phys. Rev.
Lett.} \textbf{\bibinfo{volume}{64}}, \bibinfo{pages}{2495} (%
\bibinfo{year}{1990}).

\bibitem[Gisin et~al.(2002)Gisin, Ribordy, Tittel, and Zbinden]{Gisin:2002} %
\bibinfo{author}{\bibfnamefont{N.}~\bibnamefont{Gisin}}, \bibinfo{author}{%
\bibfnamefont{G.}~\bibnamefont{Ribordy}}, \bibinfo{author}{%
\bibfnamefont{W.}~\bibnamefont{Tittel}}, and \bibinfo{author}{%
\bibfnamefont{H.}~\bibnamefont{Zbinden}}, \bibinfo{journal}{Rev. Mod. Phys.}
\textbf{\bibinfo{volume}{74}}, \bibinfo{pages}{145}
(\bibinfo{year}{2002}).

\bibitem[Marek~Zukowski(1991)]{Zukowski:1991}
\bibinfo{author}{\bibfnamefont{M.}~\bibnamefont{\ifmmode~\dot{Z}\else
  \.{Z}\fi{}ukowski}} and
\bibinfo{author}{\bibfnamefont{A.}
\bibnamefont{Zeilinger}}, \bibinfo{journal}{Phys. Lett. A} \textbf{%
\bibinfo{volume}{155}}, \bibinfo{pages}{69} (\bibinfo{year}{1991}).

\bibitem[Barreiro et~al.(2005)Barreiro, Langford, Peters, and Kwiat]%
{Barreiro:2005} \bibinfo{author}{\bibfnamefont{J.~T.} \bibnamefont{Barreiro}}%
, \bibinfo{author}{\bibfnamefont{N.~K.} \bibnamefont{Langford}}, %
\bibinfo{author}{\bibfnamefont{N.~A.} \bibnamefont{Peters}}, and %
\bibinfo{author}{\bibfnamefont{P.~G.} \bibnamefont{Kwiat}}, %
\bibinfo{journal}{Phys. Rev. Lett.} \textbf{\bibinfo{volume}{95}}, %
\bibinfo{eid}{260501} (\bibinfo{year}{2005}).

\bibitem[Boschi et~al.(1998)Boschi, Branca, De~Martini, Hardy, and Popescu]%
{Boschi:1998} \bibinfo{author}{\bibfnamefont{D.}~\bibnamefont{Boschi}}, %
\bibinfo{author}{\bibfnamefont{S.}~\bibnamefont{Branca}}, %
\bibinfo{author}{\bibfnamefont{F.}~\bibnamefont{De~Martini}}, %
\bibinfo{author}{\bibfnamefont{L.}~\bibnamefont{Hardy}}, and %
\bibinfo{author}{\bibfnamefont{S.}~\bibnamefont{Popescu}}, %
\bibinfo{journal}{Phys. Rev. Lett.} \textbf{\bibinfo{volume}{80}}, %
\bibinfo{pages}{1121} (\bibinfo{year}{1998}).

\bibitem[Michler et~al.(2000)Michler, Weinfurter, and \ifmmode~\.{Z}\else
%
\.{Z}\fi{}ukowski]{Michler:2000} \bibinfo{author}{\bibfnamefont{M.}~%
\bibnamefont{Michler}}, \bibinfo{author}{\bibfnamefont{H.}~%
\bibnamefont{Weinfurter}}, and
\bibinfo{author}{\bibfnamefont{M.}~\bibnamefont{\ifmmode~\dot{Z}\else
  \.{Z}\fi{}ukowski}}, \bibinfo{journal}{Phys. Rev. Lett.} \textbf{%
\bibinfo{volume}{84}}, \bibinfo{pages}{5457} (\bibinfo{year}{2000}).

\bibitem[Barreiro et~al.(2008)Barreiro, Wei, and Kwiat]{Barreiro:2008} %
\bibinfo{author}{\bibfnamefont{J.~T.} \bibnamefont{Barreiro}}, %
\bibinfo{author}{\bibfnamefont{T.-C.} \bibnamefont{Wei}}, and %
\bibinfo{author}{\bibfnamefont{P.~G.} \bibnamefont{Kwiat}}, %
\bibinfo{journal}{Nat. Phys.} \textbf{\bibinfo{volume}{4}}, %
\bibinfo{pages}{282} (\bibinfo{year}{2008}).

\bibitem[Hasegawa et~al.(2003)Hasegawa, Loidl, Badurek, Baron, and Rauch]%
{Hasegawa:2003} \bibinfo{author}{\bibfnamefont{Y.}~\bibnamefont{Hasegawa}}, %
\bibinfo{author}{\bibfnamefont{R.}~\bibnamefont{Loidl}}, \bibinfo{author}{%
\bibfnamefont{G.}~\bibnamefont{Badurek}}, \bibinfo{author}{%
\bibfnamefont{M.}~\bibnamefont{Baron}}, and \bibinfo{author}{%
\bibfnamefont{H.}~\bibnamefont{Rauch}}, \bibinfo{journal}{Nature} \textbf{%
\bibinfo{volume}{425}}, \bibinfo{pages}{45} (\bibinfo{year}{2003}).

\bibitem[Strekalov et~al.(1996)Strekalov, Pittman, Sergienko, Shih, and Kwiat%
]{Strekalov:1996}
\bibinfo{author}{\bibfnamefont{D.~V.}
\bibnamefont{Strekalov}},
\bibinfo{author}{\bibfnamefont{T.~B.}
\bibnamefont{Pittman}},
\bibinfo{author}{\bibfnamefont{A.~V.}
\bibnamefont{Sergienko}},
\bibinfo{author}{\bibfnamefont{Y.~H.}
\bibnamefont{Shih}}, and
\bibinfo{author}{\bibfnamefont{P.~G.}
\bibnamefont{Kwiat}}, \bibinfo{journal}{Phys. Rev. A} \textbf{%
\bibinfo{volume}{54}}, \bibinfo{pages}{R1} (\bibinfo{year}{1996}).

\bibitem[Vallone et~al.(2007)]{Vallone:2007}
\bibinfo{author}{\bibfnamefont{G.}
\bibnamefont{Vallone}},
\bibinfo{author}{\bibfnamefont{E.}
\bibnamefont{Pomarico}},
\bibinfo{author}{\bibfnamefont{P.}
\bibnamefont{Mataloni}},
\bibinfo{author}{\bibfnamefont{F.}
\bibnamefont{De Martini}}, and
\bibinfo{author}{\bibfnamefont{V.}
\bibnamefont{Berardi}}, \bibinfo{journal}{Phys. Rev. Lett} \textbf{%
\bibinfo{volume}{98}}, \bibinfo{pages}{180502} (\bibinfo{year}{2007}).

\bibitem[Chen et~al.(2007)]{Chen:2007}
\bibinfo{author}{\bibfnamefont{K.}
\bibnamefont{Chen}},
\bibinfo{author}{\bibfnamefont{C.-M.}
\bibnamefont{Li}},
\bibinfo{author}{\bibfnamefont{Q.}
\bibnamefont{Zhang}},
\bibinfo{author}{\bibfnamefont{Y.-A.}
\bibnamefont{Chen}},
\bibinfo{author}{\bibfnamefont{A.}
\bibnamefont{Goebel}},
\bibinfo{author}{\bibfnamefont{S.}
\bibnamefont{Chen}},
\bibinfo{author}{\bibfnamefont{A.}
\bibnamefont{Mair}}, and
\bibinfo{author}{\bibfnamefont{J.-W.}
\bibnamefont{Pan}}, \bibinfo{journal}{Phys. Rev. Lett} \textbf{%
\bibinfo{volume}{99}}, \bibinfo{pages}{120503} (\bibinfo{year}{2007}).

\bibitem[Raimond et~al.(1996)]{Raimond:2001}
\bibinfo{author}{\bibfnamefont{J.~M.}
\bibnamefont{Raimond}},
\bibinfo{author}{\bibfnamefont{M.}
\bibnamefont{Brune}}, and
\bibinfo{author}{\bibfnamefont{S.}
\bibnamefont{Haroche}},
\bibinfo{journal}{Rev. Mod. Phys.} \textbf{%
\bibinfo{volume}{73}}, \bibinfo{pages}{565} (\bibinfo{year}{2001}).

\bibitem[Kwiat et~al.(1995)Kwiat, Mattle, Weinfurter, Zeilinger, Sergienko,
and Shih]{Kwiat:1995}
\bibinfo{author}{\bibfnamefont{P.~G.}
\bibnamefont{Kwiat}}, \bibinfo{author}{\bibfnamefont{K.}~%
\bibnamefont{Mattle}}, \bibinfo{author}{\bibfnamefont{H.}~%
\bibnamefont{Weinfurter}}, \bibinfo{author}{\bibfnamefont{A.}~%
\bibnamefont{Zeilinger}},
\bibinfo{author}{\bibfnamefont{A.~V.}
\bibnamefont{Sergienko}}, and \bibinfo{author}{\bibfnamefont{Y.}~%
\bibnamefont{Shih}}, \bibinfo{journal}{Phys. Rev. Lett.} \textbf{%
\bibinfo{volume}{75}}, \bibinfo{pages}{4337} (\bibinfo{year}{1995}).

\bibitem[Kurtsiefer et~al.(2001)Kurtsiefer, Oberparleiter, and Weinfurter]%
{Kurtsiefer:2001} \bibinfo{author}{\bibfnamefont{C.}~%
\bibnamefont{Kurtsiefer}}, \bibinfo{author}{\bibfnamefont{M.}~%
\bibnamefont{Oberparleiter}}, and \bibinfo{author}{\bibfnamefont{H.}~%
\bibnamefont{Weinfurter}}, \bibinfo{journal}{Phys. Rev. A} \textbf{%
\bibinfo{volume}{64}}, \bibinfo{pages}{023802} (\bibinfo{year}{2001}).

\bibitem[Duan et~al.(2001)Duan, Lukin, Cirac, and Zoller]{Duan:2001} %
\bibinfo{author}{\bibfnamefont{L.~M.} \bibnamefont{Duan}}, %
\bibinfo{author}{\bibfnamefont{M.~D.} \bibnamefont{Lukin}}, %
\bibinfo{author}{\bibfnamefont{J.~I.} \bibnamefont{Cirac}}, and %
\bibinfo{author}{\bibfnamefont{P.}~\bibnamefont{Zoller}}, %
\bibinfo{journal}{Nature} \textbf{\bibinfo{volume}{414}}, %
\bibinfo{pages}{413} (\bibinfo{year}{2001}).
\end{thebibliography}
\end{document}